\documentclass[prd,showpacs,showkeys,nofootinbib,twocolumn]{revtex4-1}
\usepackage{amsmath,amsfonts,amssymb}
\usepackage{xcolor}
\usepackage[colorlinks=true,urlcolor=blue,linkcolor=blue,citecolor=blue]{hyperref}

\begin{document}

\title{Generalized nonlocal gravity framework based on Poincar\'{e} gauge theory}

\author{Dirk Puetzfeld}
\email{dirk.puetzfeld@zarm.uni-bremen.de}
\homepage{http://puetzfeld.org}
\affiliation{ZARM, University of Bremen, Am Fallturm, 28359 Bremen, Germany} 

\author{Yuri N. Obukhov}
\email{obukhov@ibrae.ac.ru}
\affiliation{Theoretical Physics Laboratory, Nuclear Safety Institute, 
Russian Academy of Sciences, B.Tulskaya 52, 115191 Moscow, Russia} 

\date{ \today}

\begin{abstract}
We describe a framework for a generalized nonlocal gravity theory inspired by Poincar\'e gauge theory. Our theory provides a unified description of previous nonlocal extensions of Einstein's theory of gravitation, in particular it allows for a clear geometrical foundation. Furthermore, it incorporates recent simplifications for the ansatz of the nonlocality, which should allow for a systematic study of the impact of nonlocal concepts on observations. 
\end{abstract}


\maketitle


\section{Introduction}\label{introduction_sec}

The study of nonlocal physical systems has a long and interesting history, both in experimental and theoretical aspects. It is instructive to recall memory-dependent phenomena (hysteresis and similar ones) in continuum mechanics, and the fundamental nonlocality in the electrodynamics of media. Moreover, the description of wave phenomena and the measurability analysis of the electromagnetic field are both necessarily nonlocal. In relativity theory, which is based on the locality postulate, the nonlocality issue arises naturally in the analysis of physical measurements made by actual observers, who operate within reference systems which are accelerating or rotating. It is obvious that an inertial system is an idealized construct, which can be realized only in a certain approximation, when the typical time and length scales of the physical process under consideration are much smaller than the characteristic time and length scales set by the acceleration and rotation of the observer's reference frame. The corresponding critical lengths $\ell_{\rm tr} = c^2/a$ and $\ell_{\rm rot} = c/\omega$ are determined by the magnitudes of acceleration $a$ and angular velocity $\omega$ of an observer, and the locality assumption is justified for physical processes on the scale $L \ll \ell_{\rm tr}$ and $L \ll \ell_{\rm rot}$. Going beyond the locality assumption, one should take into account, in an appropriate way, the average over a past world line of a noninertial observer, which results in the construction of nonlocal special relativity theory and nonlocal classical electrodynamics \cite{Mashhoon:1993,Mashhoon:2005,Mashhoon:2008}.

An extension of nonlocal special relativity to include gravitational phenomena is a highly nontrivial problem. In a nonlocal gravity (NLcG) theory, originally proposed by Mashhoon \cite{Mashhoon:2017}, gravity is assumed to be history dependent, i.e.\ the gravitational interaction has an additional feature of nonlocality in the sense of an influence (``memory'') from the past that endures. NLcG exhibits some very promising properties -- for example providing a possible solution of the dark matter problem \cite{Hehl:Mashhoon:2009:1,Blome:etal:2010:1,Chicone:Mashhoon:2012:1,Rahvar:Mashhoon:2014:1,Roshan:Rahvar:2019}. The theory is built upon an ansatz for the so-called nonlocality tensor, leading to a set of integro-differential field equations.

Based on our recent re-analysis \cite{Puetzfeld:Obukhov:Hehl:2019:1} of NLcG theory, we here suggest a nonlocal generalization inspired by the structure of Poincar\'e gauge gravity (PGT). The main motivation is as follows. From the field-theoretic point of view, NLcG is constructed as a gauge theory for the spacetime {\it translation} group $T_4$, where in the framework of the teleparallelism approach the gravitational field potential is identified with the coframe (tetrad) field $\vartheta^\alpha = e_i{}^\alpha dx^i$. The nonlocal extension is then essentially patterned after the theory of electromagnetism by replacing the Abelian gauge group $U(1)$ with $T_4$, and the potential $A = A_idx^i$ with $\vartheta^\alpha$. However, translational gauge gravity represents a degenerate case of the gauge gravity theory based on the Poincar\'e group, and it is thus natural to look for the construction of the corresponding nonlocal extension of PGT. 

Our new framework theory, termed nonlocal Poincar\'e gauge gravity (NLcPGT), draws from its conceptually clear underpinning in the form of PGT, which has been well established in a gauge gravity context over the last decades. In particular, the new framework allows for a unified description of the already known nonlocal gravity extensions in the literature. The present work can be seen as the refinement and generalization of the suggestion made in the appendix of \cite{Blome:etal:2010:1}.   

The structure of the paper is as follows: In section \ref{sec_PGT} we give a condensed account of the basic structures in Poincar\'e gauge theory. This is followed by our suggestion for the nonlocal generalization of PGT in \ref{sec_NLcPGT}. In \ref{sec_special_cases} we discuss some special limiting cases of the theory, in particular a nonlocal version of Einstein-Cartan theory. We conclude our paper in section \ref{sec_conclusions} with a discussion and outlook. An overview of our notation can be found in table \ref{tab_symbols} in appendix \ref{sec_notation}. 

\section{Poincar\'e gauge theory}\label{sec_PGT}

The gauge approach in field theory has a long history and detailed reviews of the development of the field can be found in \cite{Hehl:etal:1976,Ivanenko:1983,Blagojevic:2002,Obukhov:2006,Trautman:2006,Hehl:2013,Obukhov:2018}. The basic idea of a gauge theory of gravity based on the Poincar\'e symmetry group $G\!=\!T_4\!\rtimes\!SO(1,3)$ may be sketched as follows: The invariance of the action under an $N$-parameter group of field transformations yields, via the Noether theorem, $N$ conserved currents. When the parameters are allowed to be functions of spacetime coordinates, one needs to introduce $N$ gauge fields, which are coupled to the Noether currents, to preserve the invariance under the local (gauge) symmetry. In accordance with the general Yang-Mills-Utiyama-Kibble scheme, the 10-parameter Poincar\'e group gives rise to the 10-plet of the gauge potentials which are identified with the coframe $\vartheta^\alpha = e_i{}^\alpha dx^i$ (4 one-form potentials corresponding to the translation subgroup $T_4$) and the local connection $\Gamma^{\alpha\beta} = -\,\Gamma^{\beta\alpha} = \Gamma_{i}{}^{\alpha\beta} dx^i$ (6 one-form potentials for the Lorentz subgroup $SO(1,3)$).

Compared to general relativity (GR), in which the gravitational field equations are second-order local partial differential equations (PDEs), in the gauge approach to gravity the gravitational field equations take the form of first-order local PDEs \cite{Schouten:1989,vdHeyde:1975,Hartley:1995,Erice:1995,Nester:2010}. One can then extend the first-order {\it local} field equations to {\it nonlocal} ones via the introduction of a ``constitutive'' kernel as in the phenomenological electrodynamics of media \cite{Hehl:Obukhov:2003}. The corresponding nonlocal generalization of Einstein's theory of gravitation based on the teleparallel equivalent of GR was recently developed in \cite{Hehl:Mashhoon:2009:1,Hehl:Mashhoon:2009:2,Mashhoon:Hehl:2019}. The gauge theory of spacetime translations represents a degenerate subcase of the gauge theory of the Poincar\'e group. Here we propose a consistent generalization of the nonlocal translational gauge theory to the PGT. 

The gravitational gauge field Lagrangian density ${\cal L}_{\text{grav}}={\cal L}_{\text{grav}}(e_i{}^\alpha,T_{ij}{}^\alpha,R_{ij}{}^{\alpha\beta})$ of the underlying Riemann-Cartan spacetime is a function of the coframe $e_i{}^\alpha$, the torsion 
\begin{equation}
T_{ij}{}^\alpha := \partial_ie_j{}^\alpha - \partial_je_i{}^\alpha + \Gamma_{i\beta}{}^\alpha e_j{}^\beta - \Gamma_{j\beta}{}^\alpha e_i{}^\beta,\label{Tor}
\end{equation}
and the curvature 
\begin{equation}\label{Curv}
R_{ij}{}^{\alpha\beta} := \partial_i\Gamma_j{}^{\alpha\beta} - \partial_j\Gamma_i{}^{\alpha\beta} + \Gamma_{i\gamma}{}^\beta\Gamma_j{}^{\alpha\gamma} - \Gamma_{j\gamma}{}^\beta\Gamma_i{}^{\alpha\gamma}\,.
\end{equation}
The matter Lagrangian ${\cal L}_{\text{mat}}$ depends on the matter field(s) $\Psi$ which are minimally coupled to gravity. Then the total Lagrangian density reads
\begin{equation}
{\cal L}_{\text{tot}}={\cal L}_{\text{grav}}(e_i{}^\alpha,T_{ij}{}^\alpha,R_{ij}{}^{\alpha\beta}) + {\cal L}_{\text{mat}}(e_i{}^\alpha,\Psi,D_i\Psi)\,. \label{totLagr}
\end{equation}
Defining the two gravitational field excitations 
\begin{align}
{\cal H}^{ij}{}_\alpha&:=-2\frac{\partial{\cal L}_{\text{grav}} }{\partial T_{ij}{}^\alpha},\label{excitT}\\ 
{\cal H}^{ij}{}_{\alpha\beta}&:=-2\frac{\partial {\cal L}_{\text{grav}} }{\partial R_{ij}{}^{\alpha\beta}}\,,\label{excitR}
\end{align}
we derive the two field equations from the variation of ${\cal L}_{\text{tot}}$ with respect to $e_i{}^\alpha$ and $\Gamma_i{}^{\alpha\beta}$
\begin{align}
D_j{\cal H}^{ij}{}_\alpha-{\cal E}_\alpha{}^i&={\mathfrak T}_\alpha{}^i\,, \label{first}\\
D_j{\cal H}^{ij}{}_{\alpha\beta}-e^j{}_{[\alpha}{\cal H}^i{}_{|j|\beta]}&={\mathfrak S}_{\alpha\beta}{}^i\,,\label{second}
\end{align}
where ${\mathfrak T}_\alpha{}^i:=\delta{\cal L}_{\text{mat}}/\delta e_i{}^\alpha$ denotes the canonical energy-momentum tensor density of the matter field and ${\mathfrak S}_{\alpha\beta}{}^i:=\delta{\cal L}_{\text{mat}}/\delta\Gamma_i{}^{\alpha\beta}=-{\mathfrak S}_{\beta\alpha}{}^i$ denotes the corresponding canonical spin (angular momentum) tensor density (note that these definitions differ slightly from the ones in \cite{Hehl:etal:1980}). 

The energy-momentum tensor of the gravitational gauge fields can be expressed as
\begin{equation}
{\cal E}_\alpha{}^i=e^i{}_\alpha {\cal L}_{\text{grav}}-{\cal H}^{jk}{}_\alpha T_{jk}{}^i -{\cal H}^{jk}{}_{\alpha\beta}R_{jk}{}^{i\beta}\,.\label{gravEnergy}
\end{equation}

Equations (\ref{totLagr})-(\ref{gravEnergy}) represent the general framework for PGT, and particular gravitational models are specified by the explicit form of the gauge Lagrangian. In accordance with the general scheme of a Yang-Mills theory, we assume that the Lagrangian is local and quadratic in the Poincar\'e gauge field strengths -- torsion and curvature. The torsion tensor can be decomposed into the three irreducible pieces which we denote by $^{(I)}T_{ij}{}^\alpha$, $I=1,2,3$, whereas the curvature tensor's six irreducible pieces are denoted by $^{(K)}\!R_{ij}{}^{\alpha\beta}$, $K=1,2,...,6$ (for more details see \cite{Hehl:etal:1980,Erice:1995}). Then the quadratic PG Lagrangian reads \cite{Obukhov:2018}
\begin{align}
\stackrel{\hspace{-10pt}\text{loc}}{{\cal L}_{\text{grav}}}&=\frac{\sqrt{-g}}{2\kappa c}\Big[\left(a_0e^i{}_\alpha e^j{}_\beta - \overline{a}_0\eta^{ij}{}_{\alpha\beta}/2\right)R_{ij}{}^{\alpha\beta} - 2\lambda_0\nonumber \\  
& -\,{\frac 12}T^{ij}{}_\alpha\sum_{I=1}^3 \left(a_I\,^{(I)}T_{ij}{}^\alpha - \overline{a}_I\,^{(I)}{}{\stackrel *T}{}_{ij}{}^\alpha\right)\Big] \nonumber\\
& -\,\frac{\sqrt{-g}}{4\rho}R^{ij}{}_{\alpha\beta}\sum_{K=1}^6 \left(b_K\,^{(K)}\!R_{ij}{}^{\alpha\beta} - \overline{b}_K\,^{(K)}{\!}{\stackrel *R}{}_{ij}{}^{\alpha\beta}\right),\label{quadrLagr}
\end{align}
where $\kappa = 8\pi G/c^4$ is Einstein's gravitational constant, $\lambda_0$ is the cosmological constant, and $\rho$ is the coupling constant of ``strong gravity'' with dimension $[1/\rho] = [\hbar]$, which is mediated via the propagating Lorentz connection.  The constants $a_I, \overline{a}_I$ and $b_K, \overline{b}_K$ are dimensionless and should be of order unity. Note that we put $\overline{a}_2 = \overline{a}_3$, $\overline{b}_2 = \overline{b}_4$ and $\overline{b}_3 = \overline{b}_6$ because some of the quadratic contractions are the same. This most general quadratic Poincar\'e gravity model encompasses both the parity even and parity odd terms. The corresponding parity odd coupling constants are denoted by the overbars, whereas the dualization of the tensors is denoted by a star: $^{(I)}{}{\stackrel *T}{}_{ij}{}^\alpha =  {\frac 12}\eta_{ijkl}\,{}^{(I)}T^{kl\alpha}$ and $^{(K)}{\!}{\stackrel *R}{}_{ij}{}^{\alpha\beta} =  {\frac 12}\eta_{ijkl}\,{}^{(K)}\!R^{kl\alpha\beta}$.

We compute the excitations from the gravitational field Lagrangian (\ref{quadrLagr}) by partial differentiation according to the definitions (\ref{excitT}) and (\ref{excitR}):
\begin{align}
{\cal H}^{ij}{}_\alpha&=\frac{\sqrt{-g}}{\kappa c} \sum_{I=1}^3 \left(a_I\,^{(I)}T^{ij}{}_\alpha - \overline{a}_I\,^{(I)}{}{\stackrel *T}{}^{ij}{}_\alpha\right),\label{excitTr}\\ 
{\cal H}^{ij}{}_{\alpha\beta}&= -\,\frac{\sqrt{-g}}{\kappa c}\left(a_0 e^i{}_{[\alpha} e^j{}_{\beta]} - \overline{a}_0\eta^{ij}{}_{\alpha\beta}/2\right)\nonumber \\ 
& +\,\frac{\sqrt{-g}}{\rho}\sum_{K=1}^6 \left(b_K\,^{(K)}\!R^{ij}{}_{\alpha\beta} - \overline{b}_K\,^{(K)}{\!}{\stackrel *R}{}^{ij}{}_{\alpha\beta}\right) \nonumber \\ 
&= \,\stackrel{\text{lin}}{{\cal H}}{}^{ij}{}_{\alpha\beta}\,+\stackrel{\text{qu}}{{\cal H}}{}^{ij}{}_{\alpha\beta}\,.\label{excitRot}
\end{align}
This is the quadratic {\it local} Poincar\'e gauge theory.

\section{Nonlocal Poincar\'e gauge theory}\label{sec_NLcPGT}

In the nonlocal formulation of PGT we will make use of the bitensor formalism \cite{Synge:1960,DeWitt:Brehme:1960,Poisson:etal:2011}, in particular we adhere to the conventions of \cite{Puetzfeld:Obukhov:Hehl:2019:1}. It is worthwhile to mention that we will use a condensed notation (common to the theory of bitensors) in which the point to which the index of a bitensor belongs can be directly read from the index itself; e.g., $y_n$ denotes indices at the spacetime point $y$. Moreover, in order to distinguish the local frame indices, we use $\xi_1, \xi_2, \dots$ and $\upsilon_1, \upsilon_2, \dots$ to designate objects with frame indices at the point $x$ or $y$, in complete analogy to the labels $x_1, x_2, \dots$ and $y_1, y_2, \dots$ used in the holonomic case. 

We now generalize the local ``constitutive relations'' (\ref{excitTr}) and (\ref{excitRot}) to nonlocal ones by using an unknown scalar kernel ${\mathcal K}(x,y)$ and the parallel propagator $g_x{}^y$ for transporting tensors from point $x$ to $y$:
\begin{align}
{\cal H}^{y_1 y_2}{}_{\upsilon_3}&=\frac{1}{\kappa c} \sum_{I=1}^3 \int d^4x\sqrt{-g(x)}\,g^{y_1 x_1} g^{y_2 x_2} g_{\upsilon_3 \xi_3}\nonumber\\
\times {\mathcal K}(x,y) &\left(a_I \,^{(I)}T_{x_1 x_2}{}^{\xi_3} - \overline{a}_I \,^{(I)}{}{\stackrel *T}{}_{x_1 x_2}{}^{\xi_3} \right), \label{excitNonl1}\\ 
\stackrel{\text{lin}}{{\cal H}}{}^{y_1 y_2}{}_{\upsilon_3 \upsilon_4}&= -\,\frac{1}{\kappa c} \int d^4x\sqrt{-g(x)}\, g^{y_1}{}_{x_1} g^{y_2}{}_{x_2} g_{\upsilon_3}{}^{\xi_3}g_{\upsilon_4}{}^{\xi_4} \nonumber \\
\times {\mathcal K}(x,y) &\left( a_0 e^{x_1}{}_{[\xi_3} e^{x_2}{}_{\xi_4]} - \overline{a}_0\eta^{x_1x_2}{}_{\xi_3\xi_4}/2 \right),\label{excitNonl2}\\
\stackrel{\text{qu}}{{\cal H}}{}^{y_1 y_2}{}_{\upsilon_3 \upsilon_4}&=\frac{1}{\rho} \sum_{K=1}^6\int d^4x\sqrt{-g(x)}\,g^{y_1 x_1} g^{y_2 x_2} g_{\upsilon_3 \xi_3}g_{\upsilon_4 \xi_4}\nonumber\\
\times {\mathcal K}(x,y) &\left(b_K\,^{(K)}\!R_{x_1 x_2}{}^{\xi_3 \xi_4} - \overline{b}_K{}^{(K)}\!{\stackrel *R}{}_{x_1 x_2}{}^{\xi_3 \xi_4}\right),\label{excitNonl3}\\ 
{\cal H}^{y_1 y_2}{}_{\upsilon_3 \upsilon_4} &= \,\stackrel{\text{lin}}{{\cal H}}{}^{y_1 y_1}{}_{\upsilon_3 \upsilon_4}+ \stackrel{\text{qu}}{{\cal H}}{}^{y_1 y_2}{}_{\upsilon_3 \upsilon_4}. \label{excitNonl4}
\end{align}
This nonlocal ansatz (\ref{excitNonl1})--(\ref{excitNonl4}) should be used in (\ref{gravEnergy}) and in the field equations (\ref{first}) and (\ref{second}). In this way, we have a set of 16 + 24 integro-differential equations in terms of the variables $e_i{}^\alpha,\Gamma_i{}^{\alpha\beta}$ and $\Psi$.

\section{Special cases}\label{sec_special_cases}

\subsection{Nonlocal Einstein-Cartan theory}\label{sec_special_cases_ECT}

The torsion and the curvature square terms are absent when the coupling constants $a_I = 0$, $b_K = 0$, $\overline{a}_I = 0$, and $\overline{b}_K = 0$ vanish; in this case one recovers the Einstein-Cartan-(Holst) model which is characterized by the parity even $a_0$ and parity odd $\overline{a}_0$ coupling constants. Conventionally, one puts $a_0 = 1$ and $\overline{a}_0 = 1/\xi$,  where $\xi$ is called a Barbero-Immirzzi parameter \cite{Hojman:etal:1980,Holst:1996}. 

In the absence of matter sources with spin, the resulting nonlocal Einstein-Cartan-(Holst) theory is described by the constitutive relation (\ref{excitNonl2}) and (\ref{excitNonl4}). It represents a version of Hehl-Mashhoon nonlocal gravity theory with similar physical properties.

\subsection{Nonlocal teleparallel theory}\label{sec_special_cases_MH}

We can recover the original nonlocal teleparallel gravity theory NLcG \cite{Mashhoon:2017} when the distant parallelism condition $R_{ij}{}^{\alpha\beta} = 0$ is assumed. Then only the second line in the gravitational Lagrangian (\ref{quadrLagr}) is nontrivial, and we arrive at a generalized nonlocal teleparallel gravity with an account of parity odd terms. 

Strictly speaking, the corresponding nonlocal constitutive law (\ref{excitNonl1}) implements the improvements from \cite{Puetzfeld:Obukhov:Hehl:2019:1}, i.e.\ it avoids the unjustified complexity of the original ansatz for the nonlocality in \cite{Mashhoon:2017}, and at the same time it maintains full compatibility at the lowest orders. In particular, our choice of the nonlocality is much more natural from the viewpoint of relativistic multipolar schemes \cite{Puetzfeld:Obukhov:2014:2,Obukhov:Puetzfeld:2015:1}, since it avoids the emergence of derivatives of the world function as in \cite{Hehl:Mashhoon:2009:2}, which do not have a straightforward interpretation. Furthermore, by making use of the parallel propagator in the new ansatz (\ref{excitNonl1}), one can expect that our new constitutive law would eventually lead to the possibility of deriving exact solutions in the framework of NLcG.

\section{Discussion and conclusions}\label{sec_conclusions}

We presented a new nonlocal version of Poincar\'e gauge theory. The theory can be thought of as the generalization of the recently simplified version of NLcG discussed in \cite{Puetzfeld:Obukhov:Hehl:2019:1}. In this theory, we replace the spacetime translation gauge group $T_4$ with the Poincar\'e group $G\!=\!T_4\!\rtimes\!SO(1,3)$ and the nonlocal extension of PGT is then constructed by specifying the appropriate constitutive relation for the corresponding Poincar\'e gauge field strengths and the gravitational excitations.

The resulting model incorporates all previously known versions of Mashhoon's original theory from \cite{Mashhoon:2017}, and can be viewed as a unified, and at the same time simplified version of the original theory (in the sense how the nonlocality is implemented within the theory). In particular, it specializes to the nonlocal version of subclasses of PGT like, for example, Einstein-Cartan theory.  

Concerning the physical applications of the new theory, it is worthwhile to notice that at present nonlocal gravity is still at an early stage of development, and much needs to be done in order to establish its correspondence with experiment. Recalling the success of NLcG in giving a qualitatively and quantitatively satisfactory demonstration that the dark matter problem can be effectively addressed by a nonlocality that mimics the dark matter, we can expect a possible further elaboration of this achievement, taking into account that PGT naturally deals with the more complicated matter with microstructure \cite{Puetzfeld:Obukhov:2014:2}.

It should be noticed that so far no exact solutions of NLcPGT or NLcG are known except for the Minkowski spacetime background. However, the new simplified nonlocal ansatz for the constitutive tensor proposed here avoids the formidable technical issues of NLcG, and thereby may pave the way towards the construction of exact solutions in the full nonlinear regime of the theory. 

\begin{acknowledgments}
We are grateful to Bahram Mashhoon (Tehran) and to Friedrich W.\ Hehl (Cologne) for helpful remarks in the context of nonlocal gravity. This work was supported by the Deutsche Forschungsgemeinschaft (DFG) through the grant PU 461/1-2  –  project number 369402949 (for D.P.) and for Y.N.O. partial support was provided by the Russian Foundation for Basic Research (Grant No. 18-02-40056-mega).
\end{acknowledgments}

\appendix

\section{Notations and conventions}\label{sec_notation}
Table \ref{tab_symbols} contains a brief overview over the symbols used throughout the work.

\begin{table}
\caption{\label{tab_symbols}Directory of symbols.}
\begin{ruledtabular}
\begin{tabular}{ll}
Symbol & Explanation\\
\hline
&\\
\hline
\multicolumn{2}{l}{{Geometrical quantities}}\\
\hline
$g_{a b}$ & Metric \\
$\Gamma_{i}{}^{\alpha\beta}$ & Lorentz connection \\
$e_i{}^\alpha$ & Coframe \\
$T_{ij}{}^\alpha$ & Torsion \\
$R_{ij}{}^{\alpha\beta}$ & Curvature \\
$\eta_{ijkl}$ & Totally antisymm.\ Levi-Civita tensor \\ 
$g_{x_0}{}^{y_0}$ & Parallel propagator \\
\hline
\multicolumn{2}{l}{{Misc}}\\
\hline
${\cal L}_{\text{grav}}$, ${\cal L}_{\text{mat}}$, ${\cal L}_{\text{tot}}$ & (Gravitational,matter,total) Lagrangian \\
$\Psi$ & Matter field(s) \\
${\cal H}^{ij}{}_\alpha$, ${\cal H}^{ij}{}_{\alpha\beta}$ & Gravitational field excitations \\
${\mathfrak T}_\alpha{}^i$ & Canonical energy-momentum of matter \\
${\mathfrak S}_{\alpha\beta}{}^i$ & Canonical spin (angular momentum) \\
${\cal E}_\alpha{}^i$ & Gauge field energy-momentum \\
${\mathcal K}(x,y)$ & Causal scalar kernel \\
$\kappa$ &  Einstein's gravitational constant \\
$\lambda_0$ & Cosmological constant \\
$\rho$ & ``Strong gravity'' coupling constant \\
$a_I, \overline{a}_I,b_K,\overline{b}_K$ & Coupling constants \\
$c$ & Vacuum speed of light \\
\hline
\multicolumn{2}{l}{{Operators}}\\
\hline
$\partial_i$ & Partial derivative \\
$D_i$  & Covariant derivative \\ 
\end{tabular}
\end{ruledtabular}
\end{table}

\bibliographystyle{unsrtnat}
\bibliography{pgtnonloc_bibliography}
\end{document}